\documentclass[submission]{eptcs}
\providecommand{\event}{Post-Proceedings of TYPES 2009}
\usepackage{amssymb,alltt,graphicx,breakurl}
\def\toolong{$\hspace{-.8cm}$}
\def\smalltt#1{\texttt{\small #1}}

\title{Stateless HOL}
\author{Freek Wiedijk
\institute{
Institute for Computing and Information Sciences \\
Radboud University Nijmegen \\
Heyendaalseweg 135, 6525 AJ Nijmegen, The Netherlands}
\email{freek@cs.ru.nl}
}
\def\titlerunning{Stateless HOL}
\def\authorrunning{Freek Wiedijk}

\begin{document}
\maketitle

\centerline{\emph{Dedicated to Roel de Vrijer, in the tradition of Automath.}}
\vspace{1em}

\begin{abstract}
\parfillskip=1em plus 1fil
We present a version of the HOL Light system that supports
undoing definitions in such a way that this does not compromise the soundness
of the logic.
In our system the code that keeps track
of the constants that have been
defined thus far has been moved out of the kernel.
This means that the kernel now is purely functional.

The changes to the system are small.
All existing HOL Light developments can be run by the stateless
system with only minor changes.

The basic principle behind the system is not to name
constants by strings, but by \emph{pairs} consisting of a string and
a \emph{definition}.
This means that the data structures for the terms
are all merged into one big graph.
OCaml -- the implementation language of the system -- can use pointer equality
to establish equality of data structures fast.
This allows the system to run at acceptable speeds.
Our system runs at about 85\% of the speed of the stateful version of HOL Light.
\end{abstract}

\section{Introduction}

\subsection{Problem}

This paper describes a modification to the kernel of John Harrison's
HOL Light \cite{har:96:3,har:00} proof assistant.
Proof assistants are the best route to
\emph{complete} reliability, both
in abstract mathematics as well as for verification of computer systems.
Among the proof assistants HOL \cite{gor:mel:93} is one of the more popular
systems (next to Coq, Isabelle, PVS and ACL2), and among the
HOL implementations HOL
Light is one of the most interesting ones.
HOL Light has both been used for extensive verification of floating
point algorithms at Intel \cite{har:99,har:06:1}, as well as for impressive formalizations in
mathematics \cite{hal:07,har:08}.
Furthermore, a quite precise model of the HOL Light kernel code
has been formally proved correct \cite{har:06}.

HOL is a direct descendant of the pioneering LCF \cite{gor:mil:wad:79} system from the seventies.
In both LCF and HOL the user is not interacting with the proof
assistant through a
system specific language, but instead interacts directly with
the interpreter of the ML language in which the system has been programmed.
In the case of HOL Light this is the OCaml \cite{ler:08} language of Xavier Leroy.

For this reason in HOL there is no one keeping track of which theorems
still are valid.
Once a statement has been presented to the user as \emph{proved} -- by giving
the user
an element of the abstract datatype `\smalltt{thm}' 
as a token of its being proven -- it unavoidably will stay available
to the user as a proved statement.

This approach has the advantage that the abstract datatypes of ML
make it easy for the system to have a small proof checking \emph{kernel}
-- also called \emph{logical core} -- with the property that if
the code of that kernel can be trusted (and it implements
a consistent logic), then it is certain that
the system will be mathematically sound, i.e., it will not be possible
to prove the statement of falsity $\bot$ in it.
However, this approach has the disadvantage that it is difficult to
\emph{undo} state-modifying actions in the system.
In particular, in the existing HOL Light system it is not possible
to change the definition of a defined constant:

\begin{alltt}\small
# let X0 = new_definition `X = 0`;;
val ( X0 ) : thm = |- X = 0
# let X1 = new_definition `X = 1`;;
Exception: Failure "new_basic_definition".
\end{alltt}

\noindent
In current practice, a user of HOL Light
who wants to modify a definition will just reload the whole formalization.
This is generally not a big problem, but can become quite slow.
Also, reloading a formalization that is not finished yet and consists
of bits and pieces that have been loaded manually, can be cumbersome.

There is a good reason that HOL does not have a way of undoing definitions.
Let us suppose it would have a function
\smalltt{undo\char`\_definition}.
Then we could have the following session:

\label{sec:problem}
\begin{alltt}\small
# {let X0 = new_definition `X = 0`;;}
val ( X0 ) : thm = {|- X = 0}
{{#}} {{{undo_definition "X";;}
val it : unit = ()}}
# {let X1 = {new_definition `X = 1`};;}
val ( X1 ) : thm = |- X = 1
# {TRANS (SYM X0) X1;;}
val it : thm = |- 0 = 1
\end{alltt}
\noindent
Of course the `theorem' \smalltt{X0} will no longer be valid
after we undo the definition of \smalltt{X} as \smalltt{0},
but there is no way for us to take away
\smalltt{X0} from the user once he or she has it.
Hence, the system with \smalltt{undo\char`\_definition} clearly
is inconsistent, as one can prove \smalltt{0} \smalltt{=} \smalltt{1} in it.

The problem is that the kernel of HOL Light keeps track of
which constants have been defined already.
It has \emph{state}.
Or, stated differently, it is not purely functional.
To be able to undo definitions of constants in the HOL system,
we will switch to a \emph{stateless} kernel for the system.

\subsection{Approach}

The approach that we will follow is quite simple.
In HOL traditionally constants are \emph{named}, i.e.,
they are identified by a string of OCaml type \smalltt{string}.
We will change this to identifying the constants by the
definition itself.\footnote{
In several systems the definition of a constant consists
of two parts:
first the constant is introduced and then the value of the constant is set.
The approach from this paper does not allow such a separation.
}
That way, we do not need a state anymore to find the properties
of the constant from the name.
Then the properties will \emph{be} the name.

Actually, in our approach we will also include a string in the name
of the constant, both for convenience and for backward compatibility.
This means that the `names' that we will use for constants will
consist of a \emph{pair} of a string and a definition.
It is essential for efficiency reasons that when comparing constants
the strings will be compared first.
For this reason we put the string as the first component of the pair.

We will now look at what this means for the data structures in memory.
Here is what the constant \smalltt{`X`} from the above example
looks like in memory in the traditional stateful HOL Light system:

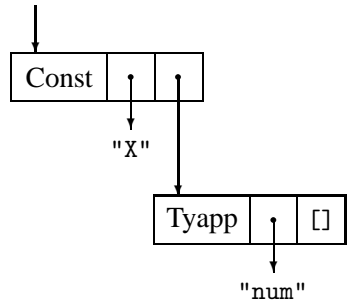
\begin{center}
\begingroup
\setlength{\unitlength}{1.8pt}
\small
\begin{picture}(135,63)(0,17)
{\put(0,60){\framebox(40,10){}}%
\put(0,60){\makebox(20,10){{\normalsize Const}}}%
\put(20,60){\line(0,1){10}}%
\put(30,60){\line(0,1){10}}}%
\put(25,65){{\circle*{1}}}
\put(35,65){{\circle*{1}}}
\thinlines
\put(5,80){{\vector(0,-1){10}}}
\put(25,65){\vector(0,-1){11}}
\put(25,65){{\circle*{1}}}
\put(15,45){\makebox(20,10){\texttt{"X"}}}
\put(35,65){{\vector(0,-1){25}}}
\put(35,65){{\circle*{1}}}
{\put(30,30){\framebox(40,10){}}%
\put(30,30){\makebox(20,10){{\normalsize Tyapp}}}%
\put(50,30){\line(0,1){10}}%
\put(60,30){\line(0,1){10}}}%
\thinlines
\put(55,35){{\circle*{1}}}
\put(55,35){\vector(0,-1){11}}
\put(55,35){{\circle*{1}}}
\put(44.8,15){\makebox(20,10){\texttt{"num"}}}
\put(60,30){\makebox(10,10){\texttt{[]}}}%
\end{picture}
\endgroup
\end{center}

\noindent
This data structure takes 40 bytes in memory (24 bytes for the two blocks
and 16 for the two strings), so this is not a large data structure.
By hiding the detail of the \smalltt{`:num`} type we can
abbreviate this as:

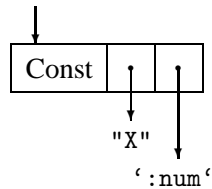
\begin{center}
\begingroup
\setlength{\unitlength}{1.8pt}
\small
\begin{picture}(135,40)(0,30)
\thinlines
\put(5,68){{\vector(0,-1){8}}}
{\put(0,50){\framebox(40,10){}}%
\put(0,50){\makebox(20,10){{\normalsize Const}}}%
\put(20,50){\line(0,1){10}}%
\put(30,50){\line(0,1){10}}}%
\thinlines
\put(25,55){{\vector(0,-1){11}}}
\put(20,35){\makebox(10,10){{\texttt{"X"}}}}
\put(35,55){{\vector(0,-1){19}}}
\put(28.8,27){\makebox(10,10){{\texttt{`:num`}}}}
\put(25,55){{\circle*{1}}}
\put(35,55){{\circle*{1}}}
\end{picture}
\endgroup
\end{center}

\noindent
Now here is the corresponding data structure for the constant \smalltt{`X`} in
our stateless HOL Light implementation:\footnote{%
We are simplifying reality slightly here.
In HOL Light the number \texttt{0} actually is the
term \texttt{`NUMERAL \char`\_0`}, so it is not a constant as drawn in the picture.
The term \texttt{\char`\_0} however \emph{is} a constant with
definition \texttt{`\char`\_0 = mk\char`\_num IND\char`\_0`}.
Tracing all further definitions from \texttt{mk\char`\_num} and \texttt{IND\char`\_0}
is an interesting exercise that we will not pursue here.}

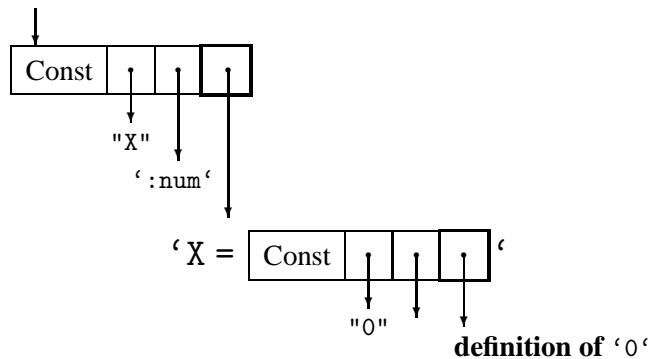
\begin{center}
\begingroup
\setlength{\unitlength}{1.8pt}
\small
\begin{picture}(135,75)(0,-5)
\thinlines
\put(5,68){{\vector(0,-1){8}}}
{\put(0,50){\framebox(50,10){}}%
\put(0,50){\makebox(20,10){{\normalsize Const}}}%
\put(20,50){\line(0,1){10}}%
\put(30,50){\line(0,1){10}}%
\put(40,50){\line(0,1){10}}}%
\thicklines
\put(40,50){\framebox(10.2,10){}}%
\thinlines
\put(25,55){{\vector(0,-1){11}}}
\put(20,35){\makebox(10,10){{\texttt{"X"}}}}
\put(35,55){{\vector(0,-1){19}}}
\put(28.8,27){\makebox(10,10){{\texttt{`:num`}}}}
\put(25,55){{\circle*{1}}}
\put(35,55){{\circle*{1}}}
\thinlines
\put(45.5,55){{\vector(0,-1){31}}}
\put(45.5,55){{\circle*{1}}}
\put(32.5,9){\makebox(77,15)[l]{{\Large\texttt{`X} \texttt{=} \hspace{6.25em} \texttt{`}}}}
{{\put(50,11){\framebox(50,10){}}}
\thicklines
\put(90,11){\framebox(10.2,10){}}%
\thinlines
\put(50,11){\makebox(20,10){{\normalsize Const}}}%
\put(70,11){\line(0,1){10}}%
\put(80,11){\line(0,1){10}}%
\put(90,11){\line(0,1){10}}%
\thinlines
\put(75,16){{\vector(0,-1){11}}}
\put(70,-4){\makebox(10,10){{\texttt{"0"}}}}
\put(85,16){{\vector(0,-1){13}}}
\put(95,16){{\vector(0,-1){15}}}
\put(75,16){{\circle*{1}}}
\put(85,16){{\circle*{1}}}
\put(95,16){{\circle*{1}}}}
\put(93,-8){\makebox(42,10)[l]{\textbf{{\normalsize definition of }\texttt{`0`}}}}
\end{picture}
\endgroup
\end{center}

\noindent
We added an extra field to the \smalltt{Const} block that points
to the definition of the \smalltt{X} constant.\footnote{%
Note that the `\texttt{X}' that occurs on the left hand side
of the defining equation is a {variable} and \emph{not}
the defined constant.
This equation was the argument to
\texttt{new\char`\_basic\char`\_definition}.

An alternative approach would be to have the new field
just point to the right hand side of the equation.
However, we considered it to be more elegant to have the field point
to the exact data that was given to the function that introduced
the constant to the system.
In this case this is the argument to the function \texttt{new\char`\_basic\char`\_definition}.
}
However, this addition is recursive:
the constant \smalltt{0} occurring in the definition of \smalltt{X}
\emph{also} has this pointer which points to the definition of \smalltt{0}.
This continues all the way until we get to pure lambda terms that
do not involve constants at all.
Only then will this chain of references end.

Clearly, the constant \smalltt{`X`} now is a large
graph consisting of blocks connected by pointers.
Of course for different constants these graphs will not be disjoint.
Therefore, all constants together should be considered to be part
of a single large data structure in memory.\footnote{
Note that although the amount of data for a single constant becomes
much larger, the amount of data for the whole system
stays roughly the same, as the data structures of the
constants will be shared.
}

Despite the size of their data structures, comparison of constants still turns out to be rather cheap.
In our modified system
different constants actually still will have different strings in their names.
Therefore, if constants are \emph{not} equal, the comparison will fail
exactly like it would fail before, i.e., while comparing the strings.
However, if the constants \emph{are} equal, their definitions in fact
also will be equal, and even will be given by equal pointers.
Now OCaml can be made to decide that two things are equal
if the pointers to them are equal.
Hence in the normal use of the HOL Light system
constant comparison will never follow the pointers to the definitions.

However, if we use the \smalltt{undo\char`\_definition} function from
the above discussion, suddenly this does not hold
anymore.
The two \smalltt{X}s, that we defined in the example will then \emph{not} be equal
(despite the fact that their strings and types are equal) because
the definitions will be different.
This then will prevent the `proof' of \smalltt{0} \smalltt{=} \smalltt{1} from working.

Our stateless system is almost identical to the stateful one.
We did not so much replace the kernel, as slightly change it.
In particular we moved the stateful part
outside the kernel.
In a picture:

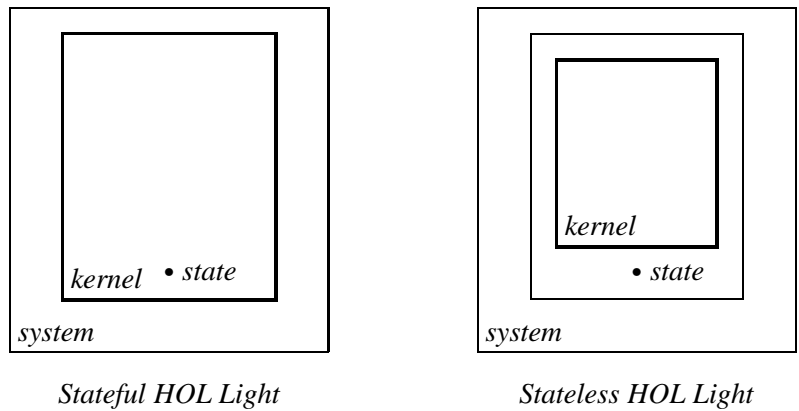
\begin{center}
\medskip
\begin{tabular}{ccc}
\begin{picture}(120,130)
\thinlines
{\put(0,0){\framebox(120,130){}}%
\put(0,0){\makebox(120,0)[lb]{\strut\ \emph{system}}}}%
\thicklines
{\put(20,20){\framebox(80,100){}}}%
{\put(20,20){\makebox(80,0)[lb]{\strut\ \emph{kernel}}}}%
\put(60,30){{\circle*{3}}}
\put(65,30){\makebox(55,0)[l]{\strut\emph{state}}}
\end{picture}%
&$\hspace{3em}$&
\begin{picture}(120,130)
{\put(0,0){\framebox(120,130){}}%
\put(0,0){\makebox(120,0)[lb]{\strut\ \emph{system}}}}%
{\put(20,20){\framebox(80,100){}}}%
\thicklines
{\put(30,40){\framebox(60,70){}}}%
\thinlines
{\put(30,40){\makebox(60,0)[lb]{\strut\ \emph{kernel}}}}%
\put(60,30){{\circle*{3}}}
\put(65,30){\makebox(55,0)[l]{\strut\emph{state}}}
\end{picture} \\
\noalign{\medskip}
\emph{Stateful {HOL} Light} && \emph{{Stateless {HOL} Light}}
\end{tabular}%
\smallskip
\end{center}

\noindent
Specifically, we split the implementation \smalltt{fusion.ml} of the kernel
(the thick box in the left picture)
into a stateless part \smalltt{core.ml} (the thick box in the right picture) and a stateful part \smalltt{state.ml} (the part outside that box),
as further described in Section~\ref{sec:code}.
This means that our system still presents a fully compatible kernel
to the rest of the system.
Therefore all existing HOL Light developments will still work with the stateless
kernel.
The only code that needs to be adapted is code that probes
the representation of constants.
This hardly every happens, and even then the changes needed are small.
We will discuss these changes in more detail below.

\subsection{Related Work}

The idea from this paper of having definitional information be part of the names of constants
is applied to type theory in \cite{geu:kre:mck:wie:10}.

Many systems use an approach similar to the one that we describe here.
For example the HOL4 system also uses a pointer internally to
distinguish different versions of the `same' constant.
However the HOL4 system currently is not purely functional.
The Matita and Epigram 1 systems use an approach similar to ours
in that they
\emph{reconstruct} the context of a term from the term itself.

Many theorem provers have a purely functional kernel.
This holds of course for all systems written in Haskell,
but for example (since version 7) also for Coq \cite{fil:00}.
However, one might argue that the kernels of these systems are not really without a notion of state,
as in those systems \emph{state} is an object that the
kernel operates on.
In Coq the state object is called an \emph{environment}, while
in HOL-based systems like ProofPower and Isabelle it is called a \emph{theory}.

In Coq, one does {not} have a separate type for well-formed terms:
to be purely functional it departs from the traditional LCF architecture in this respect.
Only a type of \emph{preterm} exists, that can be used to
\emph{extend} a state (at which time the preterm will be type checked).
Specifically in the case of Coq the basic function of the kernel essentially is:\footnote{%
In the actual Coq source code the \texttt{string} type is called \texttt{dir\char`\_path} \texttt{*} \texttt{label},
the \texttt{preterm} type is called \texttt{constr}, and
the \texttt{state} type is called \texttt{safe\char`\_environment}.}
\begin{alltt}\small
add_constant : string -> preterm -> state -> state
\end{alltt}
This function adds a new constant with a given name and expansion to the state.
Clearly the Coq kernel is functional by carrying the state
around in a monadic fashion.
In contrast, in our approach there does not exist a type corresponding to a state.

In ProofPower and Isabelle, \emph{theorems} instead of constants
are tagged, where the tag indicates which theory the theorem belongs to.
Only theorems from compatible theories are accepted by the inference rules
of those systems.
ProofPower has stateful theories, and is quite similar to HOL4.
Isabelle has a purely functional architecture, but for efficiency
reasons implements it in a non-functional way \cite{wen:02:1}.
It uses unique ids to provide an efficient
approximation of the inclusion relation on theory content -- both for
efficiency and for decidability, since a theory may contain arbitrary data
(including ML functions or fully abstract stuff).

The approaches of the various
systems show a trade-off between easy access to the implementation of the
system
(systems like HOL) and ease of navigation of the
formalizations (systems like Coq).
We show that one does not need to sacrifice
undo to get the accessibility of HOL's LCF architecture.

The trick of making the definitions part of the names of constants
occurs in logic regularly.
For instance, in Leisenring's book on the epsilon choice operator \cite{lei:69}
the term $\varepsilon x. P[x]$ takes the place of a constant name
for a witness of $\exists x. P[x]$.
That way the completeness theorem can be proved without having to
Skolemize first nor without having to add new constants.

\subsection{Contribution}

We present a version of the HOL Light system with the following
properties:
\begin{itemize}
\item
The kernel of the system is purely functional.

\item
The system supports undoing definitions in a logically sound way.

\item
The system is fully compatible with the existing HOL Light system:
all existing developments can be run with minor changes.

\item
The system runs at almost the speed of the existing HOL Light system.

\item
The kernel of the system is theoretically easier to analyze.

\end{itemize}

\subsection{Outline}

The paper is structured as follows.
In Section~\ref{sec:data} we explain how we modified the HOL data structures
and changed the kernel accordingly.
In Section~\ref{sec:code} we describe how the code of
the system further had to be modified.
In Section~\ref{sec:undo} we describe how to undo definitions.
In Section~\ref{sec:axioms} we discuss how to also have the kernel
track in a functional way which axioms have been used for which theorems.
In Section~\ref{sec:perf} we present the performance of our system.
Finally, in Section~\ref{sec:concl} we conclude.

\section{The datatypes of the logical core}\label{sec:data}

Here are the datatypes of the stateful HOL Light system (these
are the traditional HOL datatypes):

\begingroup
\def\\{\char`\\}
\def\{{\char`\{}
\def\}{\char`\}}
\begin{alltt}\small
type hol_type =
| Tyvar of string
| Tyapp of string * hol_type list\medskip
type term =
| Var of string * hol_type
| Const of string * hol_type
| Comb of term * term
| Abs of term * term\medskip
type thm =
| Sequent of term list * term
\end{alltt}
\endgroup

\noindent
Of course, these datatypes are \emph{abstract}, i.e., only
the kernel can create data of these types.
This is the essence of the LCF architecture.
In our system, we changed these definitions to the following:

\begingroup
\def\\{\char`\\}
\def\{{\char`\{}
\def\}{\char`\}}
\begin{alltt}\small
type hol_type =
| Tyvar of string
| Tyapp of hol_type_op * hol_type list\medskip
and term =
| Var of string * hol_type
| Const of string * hol_type * const_tag
| Comb of term * term
| Abs of term * term\medskip
and thm =
| Sequent of term list * term\medskip
and hol_type_op =
| Typrim of string * int
| Tydefined of string * int * \underbar{thm}\medskip
and const_tag =
| Prim
| Defined of \underbar{term}
| Mk_abstract of string * int * \underbar{thm}
| Dest_abstract of string * int * \underbar{thm}
\end{alltt}
\endgroup

\noindent
That is, we \emph{tagged} both the constants and the
defined types by the information that was used to introduce
them to the system.
In the case of the types it turned out to be more convenient
to integrate these tags with the string and arity into
a \smalltt{hol\char`\_type\char`\_op}.

The simplest example of this kind of tagging is the \smalltt{Defined}
tag for defined constants.
However the constants for the functions that map between an `abstract' defined
type and the `concrete' type from which it was carved have to be tagged
too.
(Those functions go in opposite directions: see \cite{har:00} for a detailed
description of the HOL type definition architecture.)
For this we have the \smalltt{Mk\char`\_abstract} and \smalltt{Dest\char`\_abstract} tags.

In our implementation as a sanity check we temporarily replaced the underlined types by function types,
to check that the system never followed the pointers corresponding
to these fields
(with this hack we did not have to change any other code in the system).
That is, instead of \smalltt{\underbar{term}} we used
\smalltt{unit} \smalltt{->} \smalltt{term}
and instead of \smalltt{\underbar{thm}} we used 
\smalltt{unit} \smalltt{->} \smalltt{thm}.
This works because OCaml will return \smalltt{true} if a function is compared to
itself, while it will throw the \smalltt{Invalid\char`\_argument} \smalltt{"equal:} \smalltt{functional value"} exception if it is compared to any other function.

There are various alternatives for the `tagging data' in these type
definitions.
We chose to use the exact arguments of the functions that are used
to define constants and types.
However, more economical variants are possible.
For example, instead of tagging the definition of a constant \smalltt{X}
with the defining equation \smalltt{`X} \smalltt{=} \dots\smalltt{`}, we could just
have used the body of that definition.
Similarly, in \smalltt{Tydefined}, \smalltt{Mk\char`\_abstract} and \smalltt{Dest\char`\_abstract} we could have used a \smalltt{term} instead of a \smalltt{thm}.
In that way we could have lost the dependency
of \smalltt{term} on \smalltt{thm}.
However, all these variants are really more or less equivalent,
and we chose for the one where it is most obvious where the
`tagging information' originates.

One could also consider using structure-less nonces (\smalltt{unit} \smalltt{ref}s, say) as tags.
However, this makes the system stateful again.
Also, a formal analysis of the system will be harder that way, as in that case the
tags will not be semantically meaningful.

We will now present how we adapted and reorganized the kernel according
to these new datatype definitions.
We only will do this for constants.
Analogous changes had to made for defined types.

We will start by presenting the code before we changed it.
The relevant functions for constants from the original, stateful HOL Light kernel
are:

\begingroup
\def\\{\char`\\}
\def\{{\char`\{}
\def\}{\char`\}}
\begin{alltt}\small
constants : unit -> (string * hol_type) list
definitions : unit -> thm list
get_const_type : string -> hol_type
new_constant : string * hol_type -> unit
new_basic_definition : term -> thm
mk_const : string * (hol_type * hol_type) list -> term
\end{alltt}
\endgroup

\noindent
With \smalltt{new\char`\_basic\char`\_definition} one introduces
a new constant definition to the system,
and with \smalltt{mk\char`\_const} one creates constant terms.
The implementation of these functions is completely straightforward:

\begingroup
\def\\{\char`\\}
\def\{{\char`\{}
\def\}{\char`\}}
\begin{alltt}\small
let bool_ty = Tyapp("bool",[]);;\medskip
let aty = Tyvar "A";;\medskip
let the_term_constants =
   ref ["=",Tyapp("fun",[aty;Tyapp("fun",[aty;bool_ty])])];;\medskip
let constants() = !the_term_constants;;\medskip
let the_definitions = ref ([]:thm list);;\medskip
let definitions() = !the_definitions;;\medskip
let get_const_type s = assoc s (!the_term_constants);;\medskip
let new_constant(name,ty) =
  if can get_const_type name then
    failwith ("new_constant: constant "^name^" has already been declared")\toolong
  else the_term_constants := (name,ty)::(!the_term_constants);;\medskip
let new_basic_definition tm =
  match tm with
    Comb(Comb(Const("=",_),(Var(cname,ty) as l)),r) ->
      if not(freesin [] r) then failwith "new_definition: term not closed"\toolong
      else if not (subset (type_vars_in_term r) (tyvars ty))
      then failwith "new_definition: Type variables not reflected in constant"\toolong
      else let c = new_constant(cname,ty); Const(cname,ty) in
           let dth = Sequent([],safe_mk_eq c r) in
           the_definitions := dth::(!the_definitions); dth
  | _ -> failwith "new_basic_definition";;\medskip
let mk_const(name,theta) =
  let uty = try get_const_type name with Failure _ ->
    failwith "mk_const: not a constant name" in
  Const(name,type_subst theta uty);;
\end{alltt}
\endgroup

\noindent
We will now show the corresponding code in our system.
In our stateless version of HOL Light,
the kernel interface becomes simpler.
The only relevant functions now are:

\begingroup
\def\\{\char`\\}
\def\{{\char`\{}
\def\}{\char`\}}
\begin{alltt}\small
new_prim_const : string * hol_type -> term
eq_term : hol_type -> term
new_defined_const : term -> term * thm
inst_const : term * (hol_type * hol_type) list -> term
\end{alltt}
\endgroup

\noindent
with implementation:

\begingroup
\def\\{\char`\\}
\def\{{\char`\{}
\def\}{\char`\}}
\begin{alltt}\small
let bool_tyop = Typrim("bool",0);;\medskip
let bool_ty = Tyapp(bool_tyop,[]);;\medskip
let new_prim_const(name,ty) =
  Const(name,ty,Prim);;\medskip
let eq_term ty =
  Const("=",Tyapp(Typrim("fun",2),[ty;Tyapp(Typrim("fun",2),[ty;bool_ty])]),\toolong
    Prim);;\medskip
let new_defined_const tm =
  match tm with
    Comb(Comb(Const("=",_,Prim),(Var(cname,ty) as l)),r) ->
      if not(freesin [] r) then failwith "new_definition: term not closed"\toolong
      else if not (subset (type_vars_in_term r) (tyvars ty))
      then failwith "new_definition: Type variables not reflected in constant"\toolong
      else let c = Const(cname,ty,Defined tm) in
           let dth = Sequent([],safe_mk_eq c r) in
           c,dth
  | _ -> failwith "new_basic_definition";;\medskip
let inst_const(tm,theta) =
  match tm with
  | Const(name,uty,tag) -> Const(name,type_subst theta uty,tag)
  | _ -> failwith "inst_const: not a constant";;
\end{alltt}
\endgroup

\noindent
The remainder of the code was moved out of the kernel.
These are the following functions:

\begingroup
\def\\{\char`\\}
\def\{{\char`\{}
\def\}{\char`\}}
\begin{alltt}\small
the_term_constants : (string * term) list ref
the_definitions : thm list ref
get_const_type : string -> hol_type
new_constant' : string * term -> unit
new_constant : string * hol_type -> unit
new_basic_definition : term -> thm
mk_const : string * (hol_type * hol_type) list -> term
\end{alltt}
\endgroup

\noindent
(We did not need to hide the stateful variables \smalltt{the\char`\_term\char`\_constants} and \smalltt{the\char`\_def\-ini\-tions} anymore, as
changing them can no longer compromise the logic.
For this reason we no longer need the \smalltt{constants} and
\smalltt{definitions} functions for inspecting them.)
The implementation of these functions again is straightforward:

\begingroup
\def\\{\char`\\}
\def\{{\char`\{}
\def\}{\char`\}}
\begin{alltt}\small
let the_term_constants = ref ["=",eq_term aty];;\medskip
let the_definitions = ref ([]:thm list);;\medskip
let get_const_type s = type_of (assoc s (!the_term_constants));;\medskip
let new_constant'(name,c) =
  if can get_const_type name then
    failwith ("new_constant: constant "^name^" has already been declared")\toolong
  else the_term_constants := (name,c)::(!the_term_constants);;\medskip
let new_constant(name,ty) =
  new_constant'(name,new_prim_const(name,ty));;\medskip
let new_basic_definition tm =
  let c,dth = new_defined_const tm in
  match c with
  | Const(name,_,_) ->
      new_constant'(name,c); the_definitions := dth::(!the_definitions); dth;;\toolong\medskip
let mk_const(name,theta) =
  let tm = try assoc name (!the_term_constants) with Failure _ ->
    failwith "mk_const: not a constant name" in
  inst_const(tm,theta);;
\end{alltt}
\endgroup

\noindent
Clearly, the code in our system is slightly more complicated,
but essentially it is a reorganized version of the original code.

Our code has the property that makes it easy to distinguish calls
to \smalltt{new\char`\_con\-st\-ant} that add `primitive' constants
to the system -- which in practice only is used for the Hilbert
epsilon choice operator \smalltt{`(@)`} --
from other calls to \smalltt{new\char`\_constant}.
In our system the first kind corresponds to the kernel
function \smalltt{new\char`\_prim\char`\_{\penalty 100}const}.\footnote{%
If one wants to be pedantic, one might keep track of calls to \texttt{new\char`\_prim\char`\_const} (and
to its counterpart for types) in
the \texttt{context} data structure described in Section~\ref{sec:axioms} below.
In some sense these are `axiomatic' too.}

\section{Modifications to the HOL source code}\label{sec:code}

The sizes of the kernel files are compared in the following
table:

\begin{center}
\medskip
\def\omit#1{}
\begin{tabular}{llrr}
& \emph{source file} & $\hspace{.5em}$\emph{all lines} & $\hspace{.5em}$\emph{content} \\
\noalign{\smallskip}
\strut{{\emph{Stateful {HOL} Light}}$\hspace{2em}$} \\
\noalign{\smallskip}
\hline
\noalign{\smallskip}
kernel & {\texttt{\small fusion.ml$\hspace{1em}$}} & {669} & {394} \\
\noalign{\bigskip\medskip}
\strut{{\emph{Stateless {HOL} Light}}} \\
\noalign{\smallskip}
\hline
\noalign{\smallskip}
kernel & {\texttt{\small core.ml}} & \omit{465} & {383} \\
state & {\texttt{\small state.ml}} & \omit{91} & {64} \\
\noalign{\smallskip}
\hline
\noalign{\smallskip}
{total} && \omit{556} & {447}
\end{tabular}
\end{center}

\noindent
The last column counts the number of non-blank non-comment lines.
(In our version of the code we removed all the comments, which means
that the total line count of our files is not meaningful in comparison
to the original system.)
The kernel of the stateful HOL Light is called \smalltt{fusion.ml} (it used to
be three files, \smalltt{type.ml}, \smalltt{term.ml} and \smalltt{thm.ml},
which at some point were fused).
We split this file in our kernel, called \smalltt{core.ml}, and the remainder
of the code, \smalltt{state.ml}.

In the rest of the system only two kinds of changes had to be made. 
First, the system would be impractically slow if we compared
objects using the default OCaml equality.
Recently OCaml has been changed to consider Not-A-Number floating point numbers not to be
equal to themselves, and for this reason the default equality test
never uses pointer equality as an optimization.
To get around this problem, we added the following lines at the start of
the first HOL Light source file \smalltt{lib.ml}:
\begin{alltt}\small
let (=) = fun x y -> Pervasives.compare x y = 0;;
let (<>) = fun x y -> Pervasives.compare x y <> 0;;
let (<) = fun x y -> Pervasives.compare x y < 0;;
let (<=) = fun x y -> Pervasives.compare x y <= 0;;
let (>) = fun x y -> Pervasives.compare x y > 0;;
let (>=) = fun x y -> Pervasives.compare x y >= 0;;
\end{alltt}

\noindent
Second, pattern matching on kernel datatypes had to
be changed occasionally.
As an example, in \smalltt{basics.ml} the line
\begin{alltt}\small
    Tyapp("fun",[ty1;ty2]) -> (ty1,ty2)
\end{alltt}
had to be changed to
\begin{alltt}\small
    Tyapp(Typrim("fun",2),[ty1;ty2]) -> (ty1,ty2)
\end{alltt}
\noindent
In the basic library of the system (which consists of 26,602
lines of source code) there were only 74 lines
that had to be changed like this.
These changes were systematic and could be made with
a few global replacements.

\section{Undoing definitions}\label{sec:undo}

With a stateless kernel implementing safe removal of definitions
becomes trivial.
We just add the following implementation of \smalltt{undo\char`\_definition} to the source file
\smalltt{pair.ml} (right after the implementation of \smalltt{new\char`\_definition}):

\begin{alltt}\small
let undo_definition cname =
  the_term_constants := filter ((<>) cname o fst) !the_term_constants;
  the_core_definitions := filter ((<>) cname o fst o dest_const o fst o
      strip_comb o fst o dest_eq o snd o strip_forall o concl)
    !the_core_definitions;
  the_definitions := filter ((<>) cname o fst o dest_const o fst o
      strip_comb o fst o dest_eq o snd o strip_forall o concl)
    !the_definitions;;
\end{alltt}

\noindent
This code has to be in (or after) \smalltt{pair.ml}, because only there the
variable \smalltt{the\char`_definitions} is introduced.
In fact HOL Light has \emph{two} variables with that name,
one in the kernel (in our version of course in \smalltt{state.ml} outside
the kernel), and another one in \smalltt{pair.ml}.
We renamed the first one to \smalltt{the\char`\_core\char`\_defini\-tions},
and update both variables simultaneously.

Now with this function, our motivating example session from Section~\ref{sec:problem} runs as follows:

\begin{alltt}\small
# let X0 = new_definition `X = 0`;;
val ( X0 ) : thm = |- X = 0
# undo_definition "X";;
val it : unit = ()
# let X1 = new_definition `X = 1`;;
val ( X1 ) : thm = |- X = 1
# TRANS (SYM X0) X1;;
Exception: Failure "TRANS".
\end{alltt}

\noindent
As expected the system considers the two \smalltt{X}s to  be different,
and does not allow the transitivity step anymore.

However, there still is a subtle issue.
If we now print \smalltt{X0} and \smalltt{X1},
the system will do this in the following way:

\begin{alltt}\small
# X0;;
val it : thm = |- X = 0
# X1;;
val it : thm = |- X = 1
\end{alltt}

\noindent
I.e., the system prints out what appears to be contradictory
judgements.
Of course these judgements are not \emph{actually} contradictory,
the system is perfectly sound.
The \smalltt{X} in the first \smalltt{thm} is the `old' \smalltt{X},
while the second is the `new' \smalltt{X}.
It therefore will \emph{not} be possible to prove from this that

\begin{alltt}\small
val it : thm = |- 0 = 1
\end{alltt}
\noindent
However one might ask, from a pragmatic point of view, how much
difference that makes with the confusing printout of \smalltt{X0}
and \smalltt{X1}.

This is not a problem with the consistency of the system, but
with what in \cite{wie:10} is called \emph{Pollack-consistency}.
There is nothing wrong with the kernel of the system, but with
its printing/parsing code.
The statement of theorem \smalltt{X0} is printed in a way that does \emph{not}
parse back to the same statement.
That is (using the terminology from \cite{wie:10}) the printing/parsing functions are not \emph{well-behaved}.

Of course, in \cite{wie:10} it is pointed out that the
stateful version of HOL Light already was Pollack-inconsistent.
Apparently this was not considered a serious problem, and the problem shown here
might for the same reason be ignored.
However (although we did not pursue this)
in \cite{wie:10} a simple strategy is given
to make a system Pollack-consistent, which can easily be applied here.

A simple variant of this would be to insert in the printing code
some extra lines that check whether a constant that is being printed
is equal to the `current' value of that constant, and if not to throw
an exception.
In that way it is probably easy to have the system print \smalltt{X0}
as `\smalltt{<obsolete} \smalltt{theorem>}' (or something like that)
after the definition of \smalltt{X} has been undone.
For this paper we were mainly interested in
how to minimally modify the \emph{kernel} of the system
to find out what the performance of our approach would be (and not so much
to further develop
the result into a `better' system),
therefore we have not pursued implementing this yet.

\section{Tracking the axioms}\label{sec:axioms}

The stateful HOL Light system keeps track of the axioms
that have been introduced by the user in the variable
\begin{alltt}\small
the_axioms : thm list ref
\end{alltt}
We moved this variable out of the kernel, and therefore
the system described thus far does not keep track of the
axioms that have been used for the theorems.
The whole system only uses three axioms, so one might
not consider this to be a serious problem.

However, we also investigated a variant of the system where
each \smalltt{thm} was `tagged' with the set of axioms
from which it was derived.
In that case each basic inference rule of the system had to
take the union of this set of axioms for each of the \smalltt{thm}s
that it got as arguments.
If implemented naively this would become expensive, computationally.

To get this to run with acceptable speed, we used the following
data structure\footnote{%
One of the referees of this paper pointed out that the use of the \texttt{array} type introduces
state to the kernel again, and that
this undermines the point of the paper a bit.
However, note that we use the arrays in a `purely functional' way.
We never update arrays, and only use them to be able to get to a specific index
without having to walk a list.
}.
The \smalltt{thm} type now is defined as
\begin{alltt}\small
type context =
  int * term list array\medskip
type thm =
| Sequent of context * term list * term
\end{alltt}

\noindent
The context type represents the axioms used in proving the \smalltt{thm}.
It consists of an array holding the \emph{history} of the axiom lists
during the execution of the system.
Specifically it consists of an array of lists of axioms of decreasing length,
prefixed with the length of the array minus one.
The function that introduces an axiom now is:

\begin{alltt}\small
let axiom_sequent ((n,axa) as ctx) tm =
  let axl = Array.get axa 0 in
  let ctx' = (n + 1),Array.of_list ((tm::axl)::Array.to_list axa) in
  let ax = Sequent(ctx',[],tm) in
  ax,ctx';;
\end{alltt}
Here the \smalltt{(n,axa)} argument represents the set of axioms thus far.
This is given by the stateful outside of the kernel.
The code to merge contexts is:
\begin{alltt}\small
let empty_context = 0,[|[]|];;\medskip
let merge_contexts ((n1,axa1) as ctx1) ((n2,axa2) as ctx2) =
  if ctx1 == ctx2 then ctx1 else
  if n1 < n2 then
    if Array.get axa1 0 = Array.get axa2 (n2 - n1) then ctx2 else
    failwith "merge_contexts" else
  if n1 > n2 then
    if Array.get axa1 (n1 - n2) = Array.get axa2 0 then ctx1 else
    failwith "merge_contexts" else
  failwith "merge_contexts";;
\end{alltt}

\noindent
This code, when given two contexts, does not
have to walk those contexts to see whether one is a prefix
of the other (which would cost linear time), but instead uses
the array data structure together with pointer equality
to check whether the two contexts match (taking constant time).
With this code only `compatible' contexts, where one is a subset of the
other, can be merged.
Of course a more refined version of this code, that also is able to merge sets of axioms
that are incompatible, could be written.

With this code, the `set of axioms' for theorems always will be subsets
of each other.
We call this version of the system \emph{with linear tracking of the axioms}.
We were curious whether maybe there was a theorem that, for instance,
only needed the first and third axioms but not the second one.
For this reason, we made yet another modification to the code,
that kept track of the \emph{exact} set of axioms used.
This version is called \emph{with precise tracking of axioms}.

In this version of the system we represented the set of axioms as a bit string
in a 32 bit integer.
(This version of the kernel therefore only can handle 32 axioms.
As the actual system only uses 3 axioms, for the experiment this
was sufficient.)
Now the \smalltt{context} type is:
\begin{alltt}\small
type context =
  (int * term list array) * int32
\end{alltt}
and the \smalltt{merge\char`\_context} code used OCaml's
\smalltt{Int32.logor} function on the \smalltt{int32} bitstrings.

The result of this experiment however turned out to be that for \emph{none} of the
theorems in the basic library of HOL Light that are named by a global
OCaml variable, a set of axioms is used that is not
just a prefix of the list of the three axioms in the system.
Therefore this refinement of the kernel turned out not to be very useful.

\section{Performance}\label{sec:perf}

We measured the performance of our modified HOL Light versions,
but only using wall clock time.
Specifically we used the following code in an OCaml session:
\begin{alltt}\small
#load "unix.cma";;
let starttime = ref (Unix.time());;
#use "hol.ml";;
Unix.time() -. !starttime;;
\end{alltt}
Here are the results on an unloaded Debian Etch system, using a computer with
a single $\mathop{1.86}\mathord{\mbox{GHz}}$ Intel Pentium M processor.

\begin{center}
\begin{tabular}{lrr}
\emph{version} & \emph{running time} & $\hspace{.8em}$\emph{increase} \\
\noalign{\smallskip}
\hline
\noalign{\smallskip}
Stateful & 113$\,$s \\
Stateless, without tracking of axioms & 130$\,$s & ${} + 15\%$ \\
Stateless, with linear tracking of axioms & 131$\,$s & ${} + 16\%$ \\
Stateless, with precise tracking of axioms & 132$\,$s & ${} + 17\%$ \\
\end{tabular}
\end{center}

\noindent
These are the times needed to load the basic HOL Light library.
Of course, these numbers not only represent the time spend by the
HOL Light system.
For instance, displaying the output of the system in a terminal window
already takes around 10 seconds.
Still, the table gives a reasonable impression of the performance
of the approach promoted in this paper.

\section{Conclusion}\label{sec:concl}

\subsection{Discussion}

Switching HOL Light to our stateless kernel architecture
has advantages and disadvantages:
The advantages are:
\begin{itemize}
\item
One gets the possibility to implement a function \smalltt{undo\char`\_definition}
in a logically sound way (and a similar function for type definitions).

\item
One gets a system that is probably easier to analyze theoretically.
John Harrison's HOL in HOL formalization \cite{har:06}, in which he proves
his kernel source code sound, currently leaves out the
definitions.
We expect that it will be easier to extend that work to include definitions
for our version of the system, than it would be to do this for the stateful version 
of HOL Light.

It might seem that a system with 3 non-mutual datatypes is easier
to analyze than a system with 5 mutually defined datatypes.
However, it is much less difficult to reason
about a purely functional program than about a stateful program.
This more than compensates for the slightly more involved datatypes.

In the stateful version of HOL Light the semantics of data from the kernel depend
on the state of the system, and often data has to be interpreted
in a different state than in which it was created.
This makes it hard to give the semantics in a compositional
way.
In contrast, in the stateless version all data has a direct interpretation,
making analysis much simpler.

One subtlety with proving correctness of our stateless kernel
is that it might be difficult to represent pointer equality in such a proof.
Pointer equality is rarely considered in mechanized correctness proofs of functional
programs.
However, it is clear that pointer equality is just an optimization of structural
equality and
that proving the correctness of the kernel using structural equality could be
considered equivalent.

\end{itemize}

\noindent
The disadvantages of our system compared to the stateful HOL Light are:
\begin{itemize}
\item
The system runs at about 85\% of the normal speed in daily use.

\item
The kernel is more complicated.
In particular the kernel datatype
definitions are more involved.

\end{itemize}

\noindent
We are undecided whether the slowness and added complexity in the
kernel outweighs the nicety of having a purely functional kernel
that supports undo.

\subsection{Availability}

A version of the system described in this paper can be
downloaded on the web at the web address:
\begin{alltt}\small
http://www.cs.ru.nl/~freek/notes/hol_light-stateless.tar.gz
\end{alltt}
This tar file contains just the basic library of the system,
adapted for the stateless kernel.
For reference, the tar file also contains the unmodified source code
of the HOL Light version that we used for the experiment.

\subsection{Future work}

We mainly did this experiment to satisfy our curiosity,
to find out whether the approach was viable.
We were surprised that it all worked as painlessly as it did.

We did not argue here why our changes in the implementation
are sound.
Although this seems rather obvious, it would be good to have a formal
analysis of this.
An interesting way to do this would be to adapt John Harrison's
HOL in HOL proof to also include definitions
using our stateless variant of the code.

We would like the main version of
the HOL Light system to adopt our stateless variant.
In that sense, this article can be considered an open letter
to John Harrison, asking him to consider doing this.

\paragraph{Acknowledgments.}

Thanks to Pierre Corbineau for interesting discussions on state in
the HOL Light kernel.
Thanks to Jean-Christophe Filli\^atre, Rob Arthan and Makarius Wenzel
for details of the
architecture of the Coq, ProofPower and Isabelle kernels.
Thanks to Randy Pollack for pointing out the Pollack-inconsistency
issue addressed in Section~\ref{sec:undo}.
Thanks to anonymous referees for helpful comments.


\end{document}